\documentclass{aa}  
\usepackage{graphicx}
\usepackage{natbib}
\usepackage{hyperref}
\usepackage[varg]{txfonts}
\usepackage{xcolor}
\usepackage{float}
\usepackage{dblfloatfix}  % fixes bottom placement of table* in twocolumn mode
\usepackage{afterpage}
\usepackage{placeins}
\begin{document} 

  \title{Attempting an accurate age estimate of the open cluster NGC\,6633\\ using CoRoT and Gaia}
  %\thanks{Tables X to Y are only available in electronic form at the CDS via anonymous ftp to cdsarc.u-strasbg.fr (130.79.128.5) or via http://cdsweb.u-strasbg.fr/cgi-bin/qcat?J/A+A/}}
  
  \author{K. Brogaard
          \inst{\ref{aarhus},\ref{difa}}
          \and
        A. Miglio \inst{\ref{difa},\ref{oas}}
                              \and
        T. Arentoft
                    \inst{\ref{aarhus}}
                    \and
        J. S. Thomsen\inst{\ref{difa}, \ref{aarhus},\ref{oas}} \and
                G. Casali\inst{\ref{can},\ref{difa},\ref{oas}} \and
        L. Martinelli\inst{\ref{newcastle},\ref{difa}} \and
        \\
        E. Willett\inst{\ref{bhm}} \and  
        M. Tailo\inst{\ref{difa}} 
 }

  \institute{
  {Stellar Astrophysics Centre, Department of Physics \& Astronomy, Aarhus University, Ny Munkegade 120, 8000 Aarhus C, Denmark}\label{aarhus}
  \and
  {Department of Physics \& Astronomy, University of Bologna, Via Gobetti 93/2, 40129 Bologna, Italy}\label{difa}
  \and
{INAF – Osservatorio di Astrofisica e Scienza dello Spazio, Via P. Gobetti 93/3, 40129 Bologna, Italy} \label{oas}
\and
{Research School of Astronomy and Astrophysics, The Australian National University, Canberra, ACT 2611, Australia}\label{can}
\and
{Centre for Space Science and Technology, School of Information and Physical Sciences, The University of Newcastle, University Dr, Callaghan NSW 2308, Australia}\label{newcastle}
\and
{School of Physics and Astronomy, University of Birmingham, Edgbaston, Birmingham, B15 2TT, UK}
\label{bhm}
    }

  \date{Received XXX / Accepted XXX}

  \abstract{
    % context 
   Asteroseismic investigations of solar-like oscillations in giant stars allow for the derivation of their masses and radii. For members of open clusters, this can provide an age of the cluster that should be identical to the one derived from the colour-magnitude diagram, but independent of the uncertainties that are present for that type of analysis. Thus, a more accurate age can be obtained. 
}{
    % aims
    We aim to identify and measure the properties of giant members of the open cluster NGC\,6633, and combine these with asteroseismic measurements to derive a precise and self-consistent cluster age. Importantly, we wish to constrain the effects of rotational mixing on stellar evolution, since assumptions about internal mixing can have a significant impact on stellar age estimates. 
}{
    % methods
    We identify five giant members of NGC\,6633 using photometry, proper motions, and parallaxes from Gaia, supplemented by spectroscopic literature measurements. These are combined with asteroseismic measurements from CoRoT data and compared to stellar-model isochrones. constrain the interior mixing to a low level and enabled the most precise, accurate and self-consistent age estimate so far for this cluster.
}{
    % results
    Asteroseismology, in combination with the radii of the cluster giants and the cluster colour-magnitude diagram, provides self-consistent masses of the giant members and their radii constrain the stellar interior mixing to a low level. The [C/N] ratios and Li abundances also suggest that rotation has had very little influence on the evolution of the stars in NGC\,6633. This results in an age estimate of $0.55_{-0.10}^{+0.05}$ Gyr for NGC\,6633, the most precise, accurate and self-consistent age estimate to date for this cluster.
    
    Four giant members appear to be in the helium-core burning evolutionary phase as also expected from evolutionary timescales. The bigger, cooler giant member, previously suggested to be an asymptotic giant branch star, was investigated but despite indications that the star is on the red giant branch, the evidence remains inconclusive.

}{
      % conclusion (optional)
    We derive a precise cluster age while constraining  effects of rotation and - to a lesser extent - core overshoot during the main sequence in the stellar models. A comparison to other age and mass estimates for the same stars in the literature uncovers biases for automated age estimates of helium-core burning stars. 
    }

\maketitle

\section{Introduction}

Open and globular star clusters allow investigations of stellar evolution in detail because of the properties that their member stars have in common \citep[e.g.][]{Brogaard2011,Brogaard2012,Brogaard2021b}. Asteroseismology of solar-like oscillators can further improve such studies \citep[e.g.][]{Miglio2012,Miglio2016,Handberg2017,Arentoft2019,Sandquist2020,Tailo2022,Howell2022,Brogaard2021c,Brogaard2023}. However, this requires long and uninterrupted high-precision time-series observations. Therefore, asteroseismic cluster studies are currently limited to those observed by CoRoT \citep{Baglin2006}, \textit{Kepler} \citep{Borucki2010} or K2 \citep{Howell2014}, with a few exceptions \citep[e.g. $\epsilon$ Tau analysed by][]{Arentoft2019,Brogaard2021a}. Of these missions, CoRoT only observed one open cluster, NGC\,6633 \citep{Poretti2015,Lagarde2015}. The observations allowed the derivation of average asteroseismic parameters of three cluster member giants, although with quite large uncertainties \citep{Lagarde2015}. Therefore, the masses that were derived ($2.7\pm0.6, 3.4\pm0.6$, and $4.2\pm0.9 \rm{M_{\odot}}$) and used for model comparisons by \citet{Lagarde2015} spanned a range much larger than expected for three giant stars belonging to the same cluster. Because of the large uncertainties, the numbers were still consistent within their mean mass of $3.43 \rm{M_{\odot}}$, but such a high mass of the giant stars is inconsistent with the age of NGC\,6633 in the literature, for example $\log(\rm{age})=8.84$ \citep{Cantat2020}, or $\log(\rm{age})=8.888$ \citep{Bossini2019}, which correspond to ages of 0.691 Gyr and 0.773 Gyr, respectively. For comparison, a similar open cluster, NGC\,6866, was
found to have giant masses around $2.80\, \rm{M_{\odot}}$ and an age close 0.43 Gyr \citep{Brogaard2023}. 

Young open clusters like NGC\,6633 are interesting for studies of extra mixing processes like convective-core overshooting and effects of rotation, but difficult to age-date using colour-magnitude diagrams (CMDs) due to the few and scattered stars in the turn-off region. 

A recent asteroseismic study of giant stars in the young open cluster NGC\,6866 \citep{Brogaard2023} found the amount of convective-core overshoot and rotational mixing to be significantly lower than expected, while obtaining a precise cluster age. In this work, we use similar methods in an attempt to derive an accurate cluster age for NGC\,6633, while also investigating and constraining the extra mixing processes. The paper outline is as follows. First, we present our identification of clusters members in Sect.~\ref{sec:observations} and derive luminosities for the giant members in Sect.~\ref{sec:luminosity} before gathering information on their lithium abundances and [C/N] ratios from the literature in Sect.~\ref{sec:abundances}. In Sect.~\ref{sec:asteroseismology}, we describe the asteroseismic parameters from the literature along with our own asteroseismic analysis and how we combine these results with data from Gaia DR3 \citep{GaiaDR3-2022} for more precise and robust mass and radius estimates. We then carry out comparisons to stellar-model isochrones and determine the cluster age in Sect.~\ref{sec:comparisons}. Potential consequences of our results for other areas of astrophysics are discussed in Sect.~\ref{sec:results}. Our summary, conclusions, and outlook are given in Sect.~\ref{sec:conclusions}. 

\section{Identifying targets and their properties}
\label{sec:observations}
\begin{figure}
   \centering
    \includegraphics[width=\hsize]{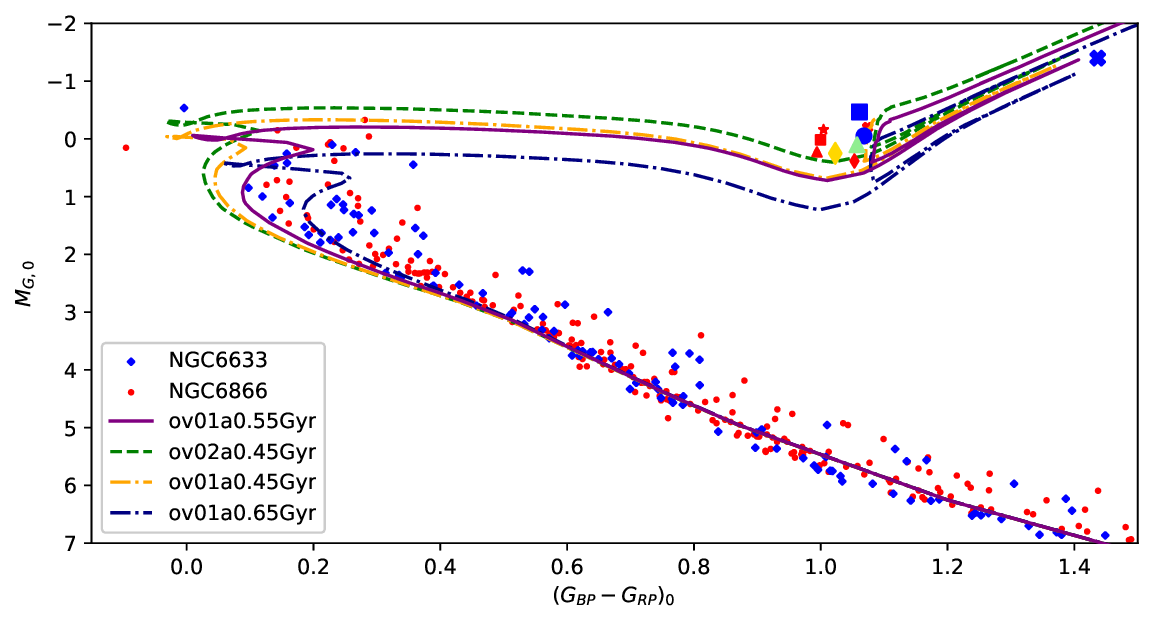}
    \includegraphics[width=\hsize]{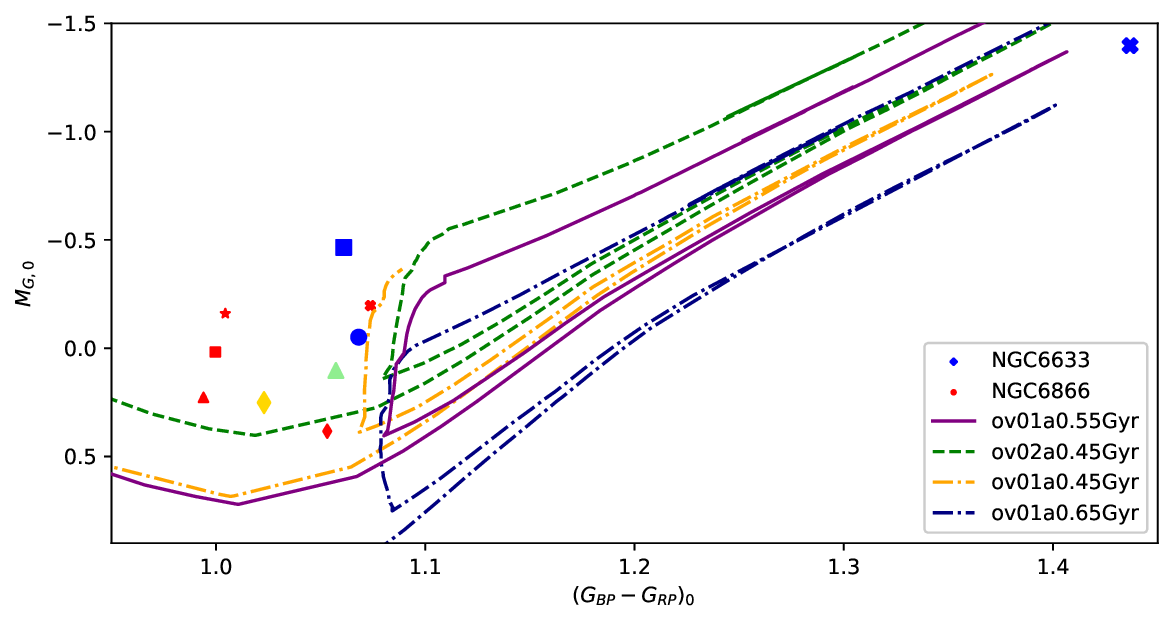}
      \caption{Gaia CMDs of NGC\,6633 and NGC\,6866 proper motion and parallax members. \textit{Top panel}: Blue diamonds mark NGC\,6633 members that have been shifted to the absolute and unreddened scale using individual Gaia parallaxes and individual reddening estimates from the 3D reddening map of \citet{Green2019}. 
      Small red circles are the members of NGC\,6866 shifted in the same way. All giant members of NGC\,6633 are marked by a specific blue, green or yellow symbol with cross reference to Table~\ref{tab:data}. The giants members of NGC\,6866 are marked with red symbols as in \citet{Brogaard2023}. Also shown are MESA isochrones with details in the legend and Sec.~\ref{sec:comparisons}. \textit{Bottom panel}: Zoom of the upper panel in the area of the giants.}
         \label{fig:CMD}
   \end{figure}

Giant members of NGC\,6633 were identified using TOPCAT \citep{Taylor2005} with the Gaia DR3 catalogue \citep{GaiaDR3-2022}. We selected stars within a 0.5 degree radius of NGC\,6633 that share similar proper motions and parallaxes:\ proper motion in right ascension from $0.514$ to 2.106 $\rm mas\,yr^{-1}$, proper motion in declination from $-2.475$ to $-1.214$ $\rm mas\,yr^{-1}$, parallax from $2.351$ to $2.722$ mas before zero-point correction. In addition, we removed stars with parallax uncertainty larger than 0.14 mas.
For the remaining stars, we derived and plotted the colour-magnitude diagram (CMD) shown in Fig.~\ref{fig:CMD} and identified the individual stars where we expected to find bright, cool cluster-member giants. We found five potential giant cluster members this way, including the three that are already established to be members with asteroseismic measurements \citep{Morel2014, Lagarde2015}. We did not do a detailed membership probability estimation, but checked instead that all five giants have a membership probability of 1.00 in the study by \citet{Hunt2023}. These giants are marked with different symbols, which are repeated in Table~\ref{tab:data} and later figures for easy cross-reference. Table~\ref{tab:data} provides an overview of the properties of these stars collected from the literature and from this work. As seen in the table, the stars have similar radial velocities, [C/N] and A(Li) abundances, providing additional evidence that they are all members of NGC\,6633. 

We derived photometric $T_{\rm eff}$ values by obtaining the reddening of each target from \citet{Green2019} and requiring $G_{\rm BP}-G_{\rm RP}$ or $G-K_S$ colours to match with the predictions from the bolometric corrections (BCs) of \citet{Casagrande2014,Casagrande2018}. The $T_{\rm eff}$ values from the two colours agree within 100 K or less for four stars and within 128 K or less for one star, which is cooler and larger than the others. There is a tendency that $G-K_S$  produces higher values of $T_{\rm eff}$ compared to $G_{\rm BP}-G_{\rm RP}$, but the general agreement is as would be expected when taking uncertainties in photometry, reddening, and colour-BC relations into account. 

We searched the literature for spectroscopic $T_{\rm eff}$ measurements of the five giants, which we list in Table~\ref{tab:data}. 
Not all targets were measured in one single study, so we made an attempt to put all five stars on the same spectroscopic $T_{\rm eff}$ scale by exploiting overlaps of targets between the different studies. 
From inter-comparisons among the spectroscopic studies, we end up adopting $T_{\rm eff}$ for HD\,170053 from \citet{Morel2014}, which was the only study that measured this star. For the other four stars, we adopt $T_{\rm eff}$ values that are about 30 K hotter than those of \citet{Casamiquela2021} to have a better agreement with the $T_{\rm eff}$ scale of \citet{Morel2014} for stars in common, while also considering star-to-star differences from the other investigations listed in Table~\ref{tab:data}.  
The agreement between the photometric and spectroscopic $T_{\rm eff}$ values is at the level of 100 K in all but a few cases, with the spectroscopic temperatures usually being higher. We adopt a $1\sigma$ uncertainty of 80 K on the $T_{\rm eff}$ values, which is larger than given in the spectroscopic studies, but likely more realistic considering the differences among studies.

All of these $T_{\rm eff}$ estimates are given in Table~\ref{tab:data} where our adopted values are marked with \textit{Ad}.

\section{Luminosities}
\label{sec:luminosity}
Luminosity estimates were derived for the five giant stars by combining the Gaia DR3 parallaxes with photometry. 
The parallaxes were zero-point corrected following \citet{Lindegren2021} with all the parameters needed taken from the Gaia archive and given in Table~\ref{tab:data} along with the derived correction. Since the stars and their sky locations are quite similar, so are the parallax corrections. The Gaia DR3 parallaxes with zero-point corrections by \citet{Lindegren2021} were shown by \citet{Khan2023} to be in very good agreement with asteroseismic predictions for stars in the \textit{Kepler} field with $G$-mag > 11, but in less good agreement at brighter magnitudes or other parts of the sky. Although the luminosities we derive could be affected by such inaccuracies, the predicted total parallax corrections for the NGC\,6633 giants are only 1.3\% of the parallaxes themselves and likely contribute a maximum error of this size.

We adopted the spectroscopic temperature estimates from the previous section along with solar metallicity and derived reddening estimates by requiring photometric $T_{\mathrm{eff}}$ values from $G-K_S$ colours and \citet{Casagrande2018} to be identical to the spectroscopic estimates. To minimise the effect of interstellar reddening and absorption uncertainties, we used 2MASS \citep{Cutri2003} $K_S$ apparent magnitudes and $K_S$ bolometric corrections from \citet{Casagrande2014} with the parallaxes to estimate the luminosities. The derived luminosities are given in Table~\ref{tab:data}. 

\section{Specific element abundances}
\label{sec:abundances}
We found abundance measurements of A(Li) for three of the giant members of NGC\,6633 in each of the works of \citet{Morel2014} and \citet{Magrini2021} with two stars in common between the samples. Four of the stars were measured by \citet{Tsantaki2023}. All sets of measurements are included in Table~\ref{tab:data}.
\citet{Morel2014} provided both LTE measurements, and values corrected for NLTE effects according to \citet{Lind2009}. \citet{Magrini2021} measured LTE values and also demonstrated the changes when adopting 3D NLTE corrections by \citet{Wang2021}. \citet{Tsantaki2023} provided LTE values also corrected to NLTE using \citet{Lind2009}.
As seen in Table~\ref{tab:data}, there is very good agreement between the A(Li) LTE values from the three studies, and the difference between the NLTE values arises from differences in the NLTE corrections by \citet{Lind2009} and the newer 3D NLTE corrections by \citet{Wang2021}. It should be appreciated that 3D is not the only difference between these two sets of NLTE corrections. For example, \citet{Wang2021} also
included a previously overlooked NLTE UV blocking effect by background opacities, which make their  abundance corrections up to 0.15 dex more negative. We refer to \citet{Wang2021} for details.

We found C and N measurements by \citet{Morel2014} for two stars and by \citet{GaiaESO1} and \citet{GaiaESO2} for three stars with one star overlapping between the two samples. The numbers are given in Table~\ref{tab:data}. Given the excellent agreement between the [C/N] for the one star in common, we assume both sets of measurements are on the same scale. 

\section{Asteroseismology}
\label{sec:asteroseismology}

The average asteroseismic parameters $\Delta\nu$ and $\nu_{\rm max}$ for HD\,170053, HD\,170147 and HD\,170231 were first derived by \citet{Lagarde2015}
using CoRoT light curves. We analysed the same data with methods described in \citet{Arentoft2017,Arentoft2019}, redetermining these parameters with higher precision and determined the observed period spacings of mixed modes $\Delta \rm{P_{obs}}$ for the first time for two of the stars. The values are given in Table~\ref{tab:seispropdata}.

\subsection{Masses}

We used the asteroseismic scaling relations as in previous works \citep[see e.g.][]{Brogaard2021c} to estimate the masses and radii of the three oscillating giants. The radius of a star is: 

\begin{eqnarray}\label{eq:03}
\frac{R}{\mathrm{R}_\odot} & = & \left(\frac{\nu _{\mathrm{max}}}{f_{\nu _{\mathrm{max}}}\nu _{\mathrm{max,}\odot}}\right) \left(\frac{\Delta \nu}{f_{\Delta \nu}\Delta \nu _{\odot}}\right)^{-2} \left(\frac{T_{\mathrm{eff}}}{T_{\mathrm{eff,}\odot}}\right)^{1/2}
\label{eq:01}
,\end{eqnarray} with the solar reference values adopted as in \citet{Handberg2017}, $\Delta \nu _{\odot}=134.9 \mu$Hz, 
$\nu _{\mathrm{max,}\odot}=3090 \mu$Hz, and $T_{\mathrm{eff,}\odot}$ = 5772 K \citep{Prsa2016}. Corrections to $\Delta \nu$, $f_{\Delta \nu}$, were determined similarly to \citet{Rodrigues2017} comparing to isochrones in diagrams for appropriate ages (e.g. see Fig. 5 in \citealt{Brogaard2023}). It was assumed that $f_{\nu _{\mathrm{max}}}=1$. Although this is not well-established from the theoretical point of view, empirical evidence from e.g. star clusters \citep{Miglio2012, Handberg2017, Brogaard2021c, Brogaard2023} and eclipsing binaries \citep{Brogaard2018, Brogaard2022} show that potential deviations from $f_{\nu _{\mathrm{max}}}=1$ are too small to be detected, at least for the ranges of masses and metallicities investigated, which encompass those in the present paper.  

Using our luminosity estimates (cf. Sect.~\ref{sec:luminosity}), we calculated an asteroseismic mass for each star in four different ways \citep{Miglio2012}:

\begin{eqnarray}
\label{eq:02}
\frac{M}{\mathrm{M}_\odot} & = & \left(\frac{\nu _{\mathrm{max}}}{f_{\nu _{\mathrm{max}}}\nu _{\mathrm{max,}\odot}}\right)^3 \left(\frac{\Delta \nu}{f_{\Delta \nu}\Delta \nu _{\odot}}\right)^{-4} \left(\frac{T_{\mathrm{eff}}}{T_{\mathrm{eff,}\odot}}\right)^{3/2},\\
\frac{M}{\mathrm{M}_\odot} & = & \left(\frac{\Delta \nu}{f_{\Delta \nu}\Delta \nu _{\odot}}\right)^{2} \left(\frac{L}{L_{\odot}}\right)^{3/2} \left(\frac{T_{\mathrm{eff}}}{T_{\mathrm{eff,}\odot}}\right)^{-6},\\
\frac{M}{\mathrm{M}_\odot} & = & \left(\frac{\nu _{\mathrm{max}}}{f_{\nu _{\mathrm{max}}}\nu _{\mathrm{max,}\odot}}\right) \left(\frac{L}{L_{\odot}}\right) \left(\frac{T_{\mathrm{eff}}}{T_{\mathrm{eff,}\odot}}\right)^{-7/2},\\
\frac{M}{\mathrm{M}_\odot} & = & \left(\frac{\nu _{\mathrm{max}}}{f_{\nu _{\mathrm{max}}}\nu _{\mathrm{max,}\odot}}\right)^{12/5} \left(\frac{\Delta \nu}{f_{\Delta \nu}\Delta \nu _{\odot}}\right)^{-14/5} \left(\frac{L}{L_{\odot}}\right)^{3/10}.
\end{eqnarray}

These mass equations are not independent, since they are just different combinations of Eqs. 1, 2, and the Stefan-Boltzmann law. Therefore, if there were no measurement errors all equations would give the same mass, and the scaling relation radius would be equal to that from the Stefan-Boltzmann law. Thus, while there are four different, but not independent, equations for the mass, there are only two equations for the radius, which is Eq. 1 and the Stefan-Boltzmann law, respectively. 

We gathered the mass and radius estimates in Table~\ref{tab:seispropdata}. The mean mass of the four equations is generally less sensitive to systematics than each individual equation because a potential larger-than-estimated error in one of the asteroseismic parameters or $T_{\rm eff}$  will have a smaller effect on the mean mass than on any of the individual equations. As seen, the mean masses of the three giants are much more similar than the previously published ones \citep{Lagarde2015}, as is expected for three giant stars belonging to the same open cluster. The difference to the previous study arises mainly because we employ the luminosity as part of the asteroseismic mass calculation, which helps minimise potential biases in the asteroseismic parameters, and decreases the uncertainties. 
In the specific case of the NGC\,6633 giants, the asteroseismic parameters have fairly large uncertainties. Therefore, the uncertainties on the masses from Eq. 4, which only uses one asteroseismic parameter, and to the lowest power, are smaller than any of the others, and the masses of the three stars are also most consistent using this equation. This equation has the additional benefit that it does not use $\Delta\nu$, which is known to be problematic in the low-$\nu_{\rm max}$ regime  \citep[e.g., see][]{Tailo2022, Zinn2023}, to which one of the giants, HD\,170053 belongs. We therefore adopt the masses from Eq. 4 and the radii from the Stefan-Boltzmann law in the following comparisons to isochrones. 

\subsection{Evolutionary states}
\label{sec:evol}

We were able to derive the so-called observed period spacing of mixed modes, $\Delta P_{\rm obs}$, for two of the three stars. The derived values can be used to distinguish between HeCB and RGB hydrogen-shell-burning phases, although sometimes not with as clear-cut a distinction as when the asymptotic period spacing, $\Delta \Pi_1$, can be measured, since the latter is the value that can be calculated from models. $\Delta P_{\rm obs}$ is always smaller than $\Delta \Pi_1$.
The HeCB phase displays higher values of period spacings but the values and differences between phases varies with stellar mass. We used the models in Figure 5 of \citet{Brogaard2023} to compare our period spacing measurements to model predictions. HD\,170174 has as a $\Delta P_{\rm obs}$ of $237\pm10 s$, which is high enough to only be compatible with the HeCB phase. HD\,170231 has $\Delta P_{\rm obs}=170\pm12 s$, which is in the $\Delta \Pi_1$ range favouring the HeCB phase but also compatible with the RGB. However, as mentioned above, $\Delta P_{\rm obs}$ is always smaller than $\Delta \Pi_1$. We used the difference between $\Delta P_{\rm obs}$ and $\Delta \Pi_1$ for a similar star in NGC\,6866, and also similar stars in NGC\,6811 \citep{Arentoft2017} for guidance on the size of this difference, see again Figure 5 of \citet{Brogaard2023}. With this, there is no doubt that HD\,170174 is in the HeCB phase.

HD\,170053 is considerably more luminous and cooler than the other NGC\,6633 giants so it is clearly not in the HeCB phase. Instead, it is either in the RGB phase or the later AGB phase. We could not measure $\Delta P_{\rm obs}$ for this star, and even if we could, it would unfortunately not allow us to distinguish between the RGB and AGB phases, since the expected period spacing at the measured $\Delta\nu$ is the same for both, as also seen in Figure 5 of \citet{Brogaard2023}.

The last two giant members were not measured with asteroseismology. To interpret them, we therefore made simulations of a 0.5 Gyr open cluster, which showed that the chance to find an RGB star is about 6\% of the chance to find a HeCB star. Demanding that the RGB star should have similar effective temperature and luminosity to that of the HeCB phase, as would be the case for these stars, reduces the number to below 1\%. This strongly suggests that the two giants without asterosemic measurements are HeCB, which is further supported by their Gaia $v_{broad}$ values being very similar to those of the confirmed HeCB stars (see Table~\ref{tab:data}); The alternative, an early first ascent RGB star in this mass range, would be rotating faster.

Summarising the above, we identified two of the giant members as clearly in the HeCB phase, another two as very likely HeCB, and one significantly more luminous star in either the RGB or AGB phase.

\section{Stellar models and isochrone comparisons}
\label{sec:comparisons}

\begin{figure}[ht!]
   \centering
    \includegraphics[width=8.4cm]{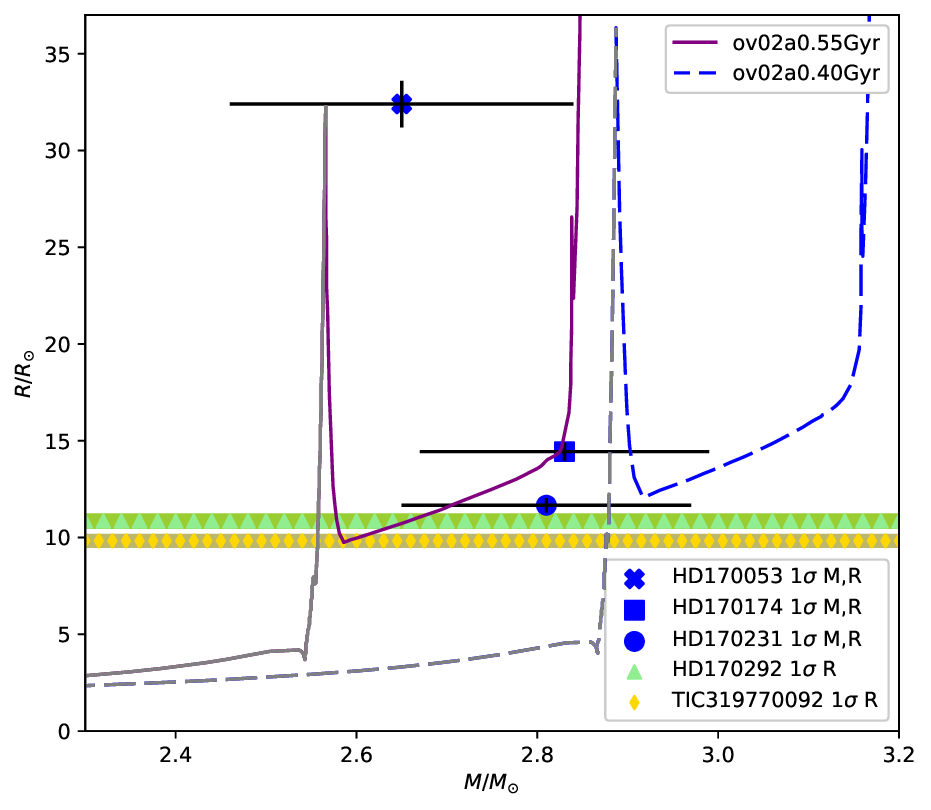}
    \includegraphics[width=8.4cm]{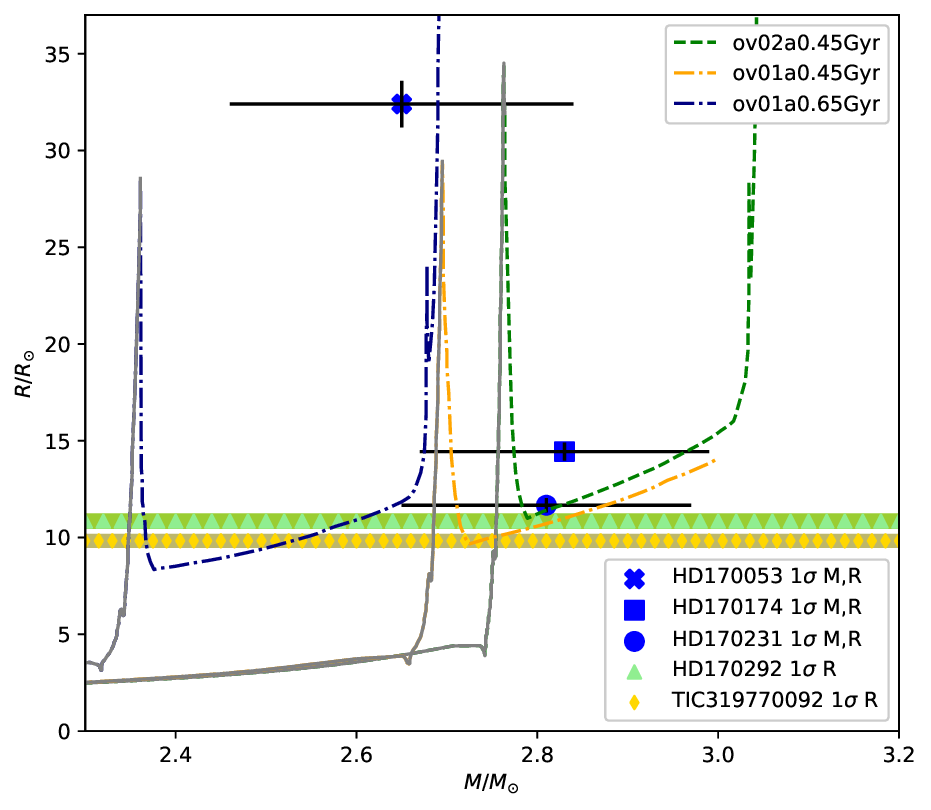}
    \includegraphics[width=8.4cm]{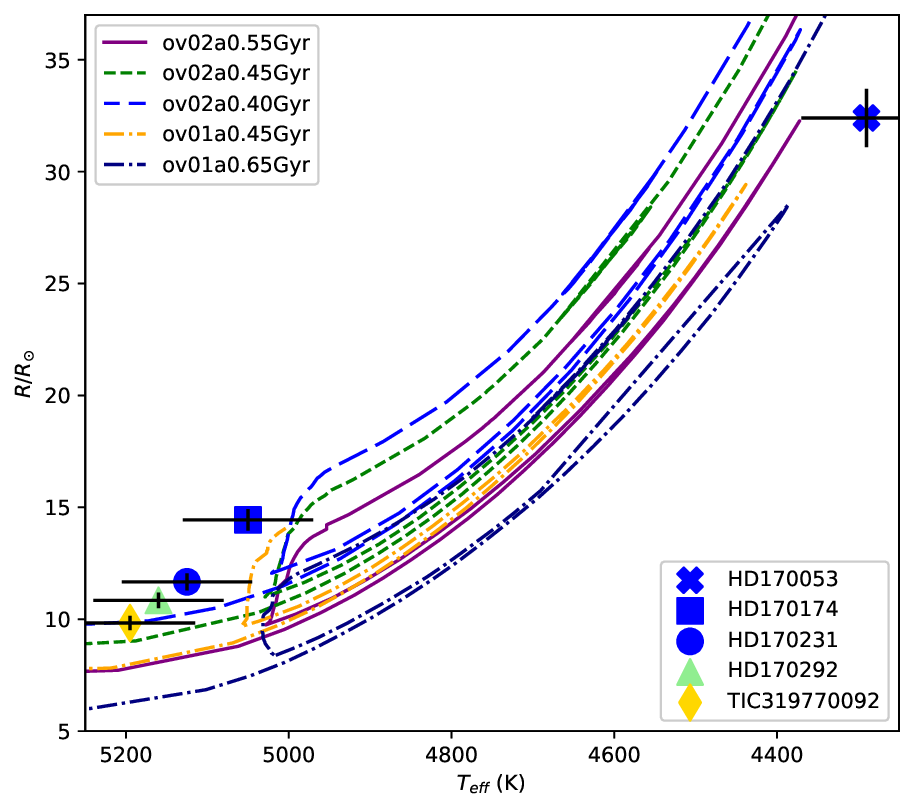}
    \caption{Mass-radius and $T_{\rm eff}$-radius diagrams for NGC\,6633 giants compared to isochrones with details in the text and legends.}
         \label{fig:mrt}
   \end{figure}

In this section, we carry out a detailed comparison of the derived properties of the NGC\,6633 giants to stellar evolution models. We first consider non-rotating models before also investigating the potential effects of rotation on the stellar evolution.

\subsection{Mass-radius diagram}

In the upper panels of  Fig.~\ref{fig:mrt}, we show the asteroseismic masses and radii of the three giants with asteroseismic measurements. The blue symbols represent the asteroseismic masses using Eq. 4 and the radii calculated from the Stefan-Boltzmann law. Thick lines with symbols represent the radii and 1$\sigma$ confidence intervals for the two giants which do not have asteroseismic measurements, HD\,17292 and TIC\,319770092. 

The measurements are compared to various isochrones calculated from MESA models \citep{Paxton2011,Paxton2013}. The details of the MESA models and isochrones are given in \citet{Rodrigues2017,Miglio2021,Campante2017,North2017,Brogaard2023}.

Several isochrones are shown in the mass-radius diagram in the upper panels of Fig.~\ref{fig:mrt} and compared to the mass and radius measurements, with the specific properties of each given in the legends. Starting on the lower left in the upper panels, the isochrones show the upper main sequence, shifting to the subgiant and red giant phases where they become almost vertical. The isochrones then reach the red giant branch (RGB) tip towards the top of the figure. For these relatively massive stars there is no helium flash but a transition into the Helium-core burning (HeCB) phase, which causes the isochrones to change direction into the fast-descending beginning of the core-helium burning phase. This ends in the stable HeCB phase, which lasts much longer and with little radius change. Therefore, the isochrones only slope upwards slightly on this part in the mass-radius diagram, before becoming almost vertical again when entering the asymptotic giant branch (AGB) phase, where for some of the isochrones the AGB bump is visible as a small wiggle. Because the HeCB phase lasts much longer than the RGB and AGB phases it most likely that all the cluster giants, with the exception of the largest one, are in the stable HeCB phase. Therefore, we have coloured only the part of the mass-radius isochrones that represent the phases that are to be compared to the data, and left the other parts grey. The largest and most luminous star is either on the AGB, in the very late RGB, or in the very early HeCB phase close to the RGB tip. We will return to this later.

Comparing the asteroseismic parameters to the isochrones in the mass-radius diagram in the upper panels of Fig.~\ref{fig:mrt} allows constraints on the cluster age. First, we consider the blue long-dashed isochrone with an age of 0.40 Gyr in the topmost panel. This represents the extreme lower limit on the cluster age given that it matches the two HeCB stars at the edge of their upper 1$\sigma$ error-bars and is outside the 1$\sigma$ error-bar of the bigger giant regardless of its evolutionary state. Furthermore, this scenario requires the two giants without asteroseismic measurements to be RGB stars because the HeCB phase does not reach low enough radii to match them. The RGB phase was however in contradiction with our findings in Sect.~\ref{sec:evol} based on evolutionary timescales and $v_{\text{broad}}$ values. Instead, the purple solid isochrone with an age of 0.55 Gyr represents the youngest age, which is consistent with all measured radii and at the same time matches the asteroseismic masses of the other stars very well. This thus represents a good age estimate.

The isochrones in the top panel of Fig.~\ref{fig:mrt} assume a core-overshooting parameter during the main-sequence $\alpha_{\rm ov} = \beta \cdot H_p$, where $H_p$ is the pressure scale height at the convective boundary and $\beta = 0.2$ \citep{Rodrigues2017}. In the middle panel, the orange dash-dotted isochrone shows how the youngest well-matching isochrone changes with the adoption of a lower value of the convective-core overshooting parameter $\beta=0.1$; all stars can be well-matched at a slightly younger age, and the best age estimate shifts downwards by about 0.10 Gyr. The green dashed isohrone with the same age of 0.45 Gyr and $\beta = 0.2$ allows a direct comparison of the effects of the overshooting efficiency.

The age-constraining powers of our measurements towards older ages depend strongly on whether the biggest star, HD\,170053, is on the RGB or the AGB. If it belongs to the RGB then the purple solid isochrone in the upper panel of Fig.~\ref{fig:mrt} is close to the upper 1$\sigma$ age limit - an older isochrone will have the RGB phase at a mass lower than the 1$\sigma$ lower limit of HD\,170053 and would also not reach large enough radii. Thus, since the models also cannot be younger without overpredicting the radius of TIC\,319770092 for the HeCB phase, they are constrained within a very small age-interval. But if HD\,170053 is instead on the AGB, then older models also match the observations within uncertainties. The navy-blue dot-dashed isochrone with an age of 0.65 Gyr still matches the asteroseismic masses of the HeCB stars at the extreme of their 1$\sigma$ error-bars. However, in this and older scenarios the giant stars are all piled up at the end of and after the HeCB phase, with no stars in the earlier part of the HeCB phase, at odds with evolutionary time-scales. Furthermore, for these older isochrones, it becomes increasingly difficult to match HD\,170174 represented by the solid square in the radius-$T_{\rm eff}$ diagram in the lower panel of Fig.~\ref{fig:mrt}. The model temperatures are offset compared to the measurements, but tension remains even if the isochrones are shifted in temperature, as discussed below. 

\subsection{Radius-$T_{\rm eff}$ diagram}
\label{sec:rt}
Whether or not HD\,170053 is an RGB or AGB stars remains uncertain. \citet{Smiljanic2009} suggested that this star could be an early-AGB star based on its position in a $V,B-V$ CMD. This was also noted by \citet{Lagarde2015}, who mentioned that this is in agreement with the carbon isotopic ratio of $18\pm8$ for the star. However, they fail to mention that the RGB phase is also in perfect agreement with this value, and therefore the carbon isotopic ratio cannot be used to discriminate, see the right panel of their Fig. 14. We show in the lower panel of Fig~\ref{fig:mrt} the radius-$T_{\rm eff}$ diagram. The isochrones run from the beginning of the RGB on the lower left to the RGB tip on the upper right, then return down to the stable HeCB phase on the lower left before returning to the upper right and out of the diagram through the AGB phase. As seen, the spectroscopic effective temperature of HD\,170053 is closer to the RGB phase than the AGB phase, while in Fig.~\ref{fig:CMD} the evolutionary phase based on colour depends on whether one trusts the 3D reddening map of \citet{Green2019} over a common reddening for all cluster members to a decisive level. The individual star reddening estimate from \citet{Green2019} is lower for HD\,170053 than the other cluster giants, and this is consistent with the reddening calculated assuming the spectroscopic $T_{\rm eff}$ values (see Table~\ref{tab:data}). Therefore, an assumption of a common reddening for all cluster stars must have been the reason that the star appears to be on the AGB in the CMD of \citet{Smiljanic2009}. 

The radius-$T_{\rm eff}$ diagram in the lower panel of Fig.~\ref{fig:mrt} demonstrates how difficult it is to extract information using the effective temperature. 
The observed mismatch between observations and models could for example be due to the assumed surface-boundary conditions and/or mixing-length parameter. We refer the interested reader to \citet{Brogaard2023}, where it is shown that including diffusion in the models increases their effective temperature in the HeCB phase (and other giant phases) due to the resultant change to the mixing length parameter in the solar calibration.
However, in our case the measurements of the HeCB stars are all hotter than the isochrones, while the measurements of HD\,170053 is cooler than the isochrones. Thus, agreement cannot be reached by applying a simple shift to either measurements or observations. Since HD\,170053 is not matched by any of the isochrones, it could be either an RGB star very close to the RGB tip or an AGB star. Interestingly, for a low core-overshoot scenario, $\beta = 0.1$ the RGB tip radius of the model does not reach the radius of the observed star, so in that scenario it would have to be an AGB star. Unfortunately, we do not have tight constraints on the overshooting, and for $\beta = 0.2$ the RGB tip reaches the radius of the measurement, although $T_{\rm eff}$ of the models is too high. The tension in $T_{\rm eff}$ is worse for the AGB scenario, but given the overall poor agreement between model temperatures and observations, $T_{\rm eff}$ seems a weak argument to prefer one evolutionary stage over another. Perhaps, in the future, asteroseismic modelling of the oscillation frequencies can shed new light on this, though it might require new higher-quality space-based time-series observations than what is available at present.  

\subsection{Colour-magnitude diagram and comparison to NGC\,6866}

The Gaia CMD of Fig.~\ref{fig:CMD} compares the cluster sequences of NGC\,6633 and NGC\,6866 when the individual distances, reddening and absorption have been removed using Gaia DR3 parallaxes and the 3D reddening map of \citet{Green2019}. The giant stars of NGC\,6633 are marked with individual symbols and the giant stars of NGC\,6866 are marked with red symbols as in the analysis by \citet{Brogaard2023}. The Gaia colours of the giant stars are similar as one would expect from the similarity between the $T_{\rm eff}$ of the giants of these two clusters. The star-to-star colour-differences of the giants are larger than expected given that the models and the spectroscopic measurements predict similar temperatures for the giant stars in both clusters. This demonstrates the limitations when trying to infer $T_{\rm eff}$ and reddening from photometry.

At the turn-off of the cluster, the stars of NGC\,6633 reach slightly bluer and less bright magnitudes than those of NGC\,6866. The difference between the bluest colour reached in the two clusters is at the level where uncertainties in reddening will affect conclusions significantly. Disregarding that, NGC\,6633 could be reaching bluer colours than NGC\,6866 for several reasons; it could be less metal-rich, have a higher degree of internal mixing, or be younger. We were not able to find a combination of parameters that would cause isochrones to simultaneously display similar differences in turn-off colour and luminosity at the top of the main-sequence as the observations of the two clusters. A difference in metallicty should also reveal itself as a colour-difference between the main-sequences, which it does not.

Fig.~\ref{fig:CMD} also compares the observed CMD to the same isochrones as used in Fig.~\ref{fig:mrt}. The minimum magnitude reached by stars at the top of the main-sequence support the conclusions reached from the mass-radius diagram of the giants. It is the magnitude of the isochrone at the reddest point on the main sequence brighter than the turn-off, just before it bends back towards the blue in the fast contraction phase, which should be compared to the minimum magnitude observed on the upper main sequence. The later evolutionary stages in the fast contraction phase and on the subgiant branch are so fast that one would not expect to observe stars there.

A 0.40 Gyr isochrone reaches brighter than the observed magnitudes at the top of the main-sequence. The same is true for the 0.45 Gyr green isochrone shown, which assumes $\beta = 0.2$. The brightest cluster stars on the main-sequence thus indicate ages larger than this, and support the previous suggestion that the least luminous of the giants, marked with a yellow diamond, is a HeCB star, since this is compatible with the isochrones for ages of 0.5 Gyr or older, see Fig.~\ref{fig:mrt}. The orange and purple isochrones both match the brightness of the minimum magnitude on the main-sequence quite well, but differ in both age and core-overshoot efficiency. The case with $\beta=0.2$
matches the turn-off colours better, but unfortunately it is difficult to know to which level this can be used as a reliable model constraint given the many issues with colours of turn-off stars in young open clusters, and the apparent tension with asteroseismology of the giant stars in NGC\,6866 \citep{Brogaard2023}, which preferred $\beta \leq 0.1$. Unfortunately, in the case of NGC\,6633, the current asteroseismic measurements are not precise enough to constrain the core-overshoot at a level to distinguish between $\beta = 0.2$ or lower.
At the older age of 0.65 Gyr, represented by the navy isochrone, there is a clear mismatch with both turn-off colour and magnitude, which is consistent with the tension already noted in the mass-radius and mass-$T_{\rm eff}$ diagrams.

\subsection{Best age estimate from non-rotating models}

Taking all age considerations together, we estimate an age of $0.55_{-0.05}^{+0.05}$ Gyr for NGC\,6633 if $\beta=0.2$ or $0.50_{-0.05}^{+0.05}$ for $\beta=0.1$ with the uncertainties likely being conservative - towards younger ages because the isochrone radii for the HeCB stage become too large to match the observations and towards older ages because all the giants pile up towards and after the end of the HeCB phase. In both cases, this is supported by how well the isochrones match the magnitude of the brightest main-sequence stars in the CMD. If the largest giant could be proven to be in the RGB phase, then smaller uncertainties could be obtained towards higher ages.
These age estimates do not include uncertainties related to uncertainty in [Fe/H], but we note that the overlap of the main-sequences of NGC\,6633 and NGC\,6866
suggest that these clusters have very similar metallicity, and literature measurements suggest that $\rm [Fe/H]=+0.0$ or very close for both clusters \citep{Santos2009,Morel2014,Casamiquela2021,Magrini2021}. The most discrepant study found $\rm [Fe/H]=-0.1$ for NGC\,6633 \citep{Jeffries2002}, which would not have a big impact on the age we derived, since our main constraint is from the minimum radius in the stable HeCB phase. The relatively large asteroseismic mass uncertainties make them secondary constraints. Therefore, in the mass-radius diagram in Fig.~\ref{fig:mrt}, an isochrone of the same age but lower metallicity is shifted slightly to the left towards lower masses without changing shape (by about 0.05 $M_{\odot}$ as inferred from second panel in Fig. 8 of \citealt{Brogaard2023}). Therefore, almost identical ages would be obtained, but for slightly lower masses of the giants.

\subsection{Models with rotation}
\label{sec:rot}

We investigate here what the abundances of lithium, the [C/N] ratios and the radii of the giant stars indicate regarding the rotational evolutionary effects in, the giant members of NGC\,6633. 

\subsubsection{Li abundance inferences}

 \begin{figure}[ht]
   \centering
    \includegraphics[width=\hsize]{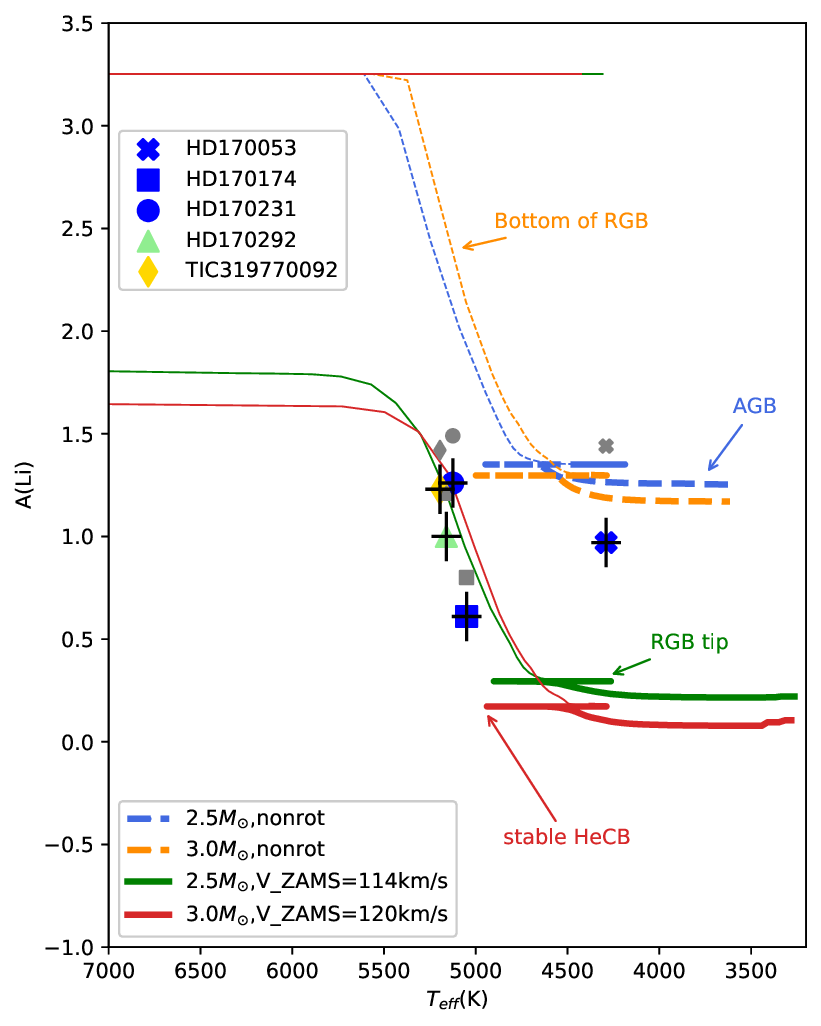}
      \caption{A(Li) measurements of NGC\,6633 giant stars as a function of $T_{\rm eff}$ compared to stellar evolutionary tracks \citep{Lagarde2012} with and without rotational effects for solar metallicity ($Z=0.014$) and relevant masses. The beginning of different evolutionary stages are marked with descriptions.
      The coloured symbols represent LTE A(Li) measurements for the giant stars of NGC\,6633 from \citet{Morel2014,Magrini2021,Tsantaki2023} with 3D NLTE corrections from \citet{Wang2021}, while the grey symbols mark the same measurements, but with NLTE corrections from \citet{Lind2009}.
      The observations should be matched by the thick horizontal parts of the tracks, and the apparent match between models and observations for the green and red tracks is not real, as it assumes all giants belong to the fast-lived RGB phase.}              \label{fig:Li}
   \end{figure}

Using the Li abundances we re-investigated Fig. 14 of \citet{Lagarde2015}, which compares measurements to models in a $T_{\rm eff}$-A(Li) plane. Our version is shown in Fig.~\ref{fig:Li}, where we have marked different evolutionary stages on the figure, which include evolutionary tracks from \citet{Lagarde2012}. Compared to \citet{Lagarde2015}, we were able to add two more measurements to this plot using the mean differences of the Li abundances of overlapping stars between \citet{Morel2014}, \citet{Magrini2021} and \citet{Tsantaki2023}. We show both the A(Li) values using the most recent 3D NLTE corrections by \citet{Tsantaki2023}, with the same coloured symbols as in previous figures, and using the previous NLTE corrections by \citet{Lind2009}, with grey symbols, to demonstrate the size of the difference. 
Keeping in mind that the measured masses of the stars are about $2.6-2.9 \rm{M_{\odot}}$ and that they are HeCB stars except for the coolest and brightest star, they should be matched by the horizontal parts of the models close to 5000 K, marked as "stable HeCB" on the red track. The apparent offset in $T_{\rm eff}$ is due to errors in either the measurements, the models, or both.
The models predict that the Li abundance should not change much during the HeCB phase. If this is true, then the star-to-star scatter in Li for the four HeCB stars could be due to observational errors, but the scatter is larger than suggested by the uncertainty estimates of 0.12 given by \citet{Morel2014}. There is no correlation between A(Li) and luminosity, which would be expected if the star-to-star difference were real and related to evolution. The Li-abundance of the coolest giant HD\,170053 does unfortunately not allow us to distinguish between the RGB and AGB evolutionary phases, since the model-predicted differences in A(Li) between these two phases is very small compared to the observational uncertainty. On the other hand, this means that we can use a mean of the Li measurements of all the giants to estimate where the horizontal parts of a well-matching model, corresponding to the parts including the RGB tip, the HeCB phase, and the AGB phase, should be. Before doing that, we stress again that it is only the horizontal part of a model track between the part marked as "stable HeCB" and "RGB tip", that should be matched to the observations, since the observed stars have an overwhelmingly larger likelihood of being HeCB stars rather than RGB stars. Thus the apparent match of the rotating models to the observations in Fig.~\ref{fig:Li} is an illusion, and the non-rotating models are much closer to a match to the observations. This was not made clear in the works of \citet{Lagarde2015} and similar analysis of Lithium in M48 in Figure 7 of \citet{Sun2023}. The mismatch in $T_{\rm eff}$ is due to errors in either the observed or model temperature scale, or even both, as discussed in Sect.~\ref{sec:rt}.

The measured Li abundances are slightly lower than the non-rotating models, so if we assume an approximately linear change in A(Li) with increasing rotation speed between $V_{\rm ZAMS} = 0$ and $V_{\rm ZAMS} = 120\,\rm km\,s^{-1}$, the stars appear to have been rotating quite slowly for their masses with a mean speed of about 40\,$\rm km\,s^{-1}$ at the ZAMS. The large difference in A(Li) between models with slightly different rotation rates means that the star-to-star scatter could be related to small differences in their initial rotation rates, which from Fig.~\ref{fig:Li} can be estimated to a total maximum full range of variation of about 70\,$\rm km\,s^{-1}$ or less around the mean of 40 $\rm km\,s^{-1}$. The problem with that interpretation is that it is inconsistent with the $v\sin i$ measurements at the turn-off of NGC\,6633; among 22 member stars with measured $v\sin i$ and $T_{\rm eff}$ > 6800 K analysed by \citet{Hourihane2023}, 6 has $v\sin i$ in the range 200-280 $\rm km\,s^{-1}$, 9 in the range 100-200\,$\rm km\,s^{-1}$ and 7 was below 100\,$\rm km\,s^{-1}$. Thus, the measured rotational velocities of the stars at the top of the main-sequence reach mean values that are much larger than suggested by the Lithium abundances of the giants. This is likely an indication that the influence of rotation on the Lithium abundance is smaller than predicted by current models.

Comparing our measurements to HeCB stars with similar masses in other open clusters measured by \citet{Magrini2021} (third vertical plot in the right-hand panels of their Fig. 11) suggests that the stars in NGC\,6633 (and NGC\,6866) are among the HeCB giants with the slowest rotation history on the main-sequence in young open clusters, when measured indirectly from their Li abundance. The same study also shows that all the young open clusters (age $\leq 707$ Myr) include HeCB stars where the lithium abundances are well-matched by a non-rotating model, while other stars in the same clusters have A(Li) that indicate some rotation. This is seen in the top row of their Fig. 12, where we recall that the observations should be compared to the horizontal parts of the tracks starting close to 5000 K. While this may just be the manifestation of the variation of velocities the stars had at the ZAMS, it still means that either the mean ZAMS velocity is quite low and the initial velocity distribution extends to close to zero even at these fairly high masses, or the effects of rotation on the lithium abundance is poorly modelled. A comparison of the Lithium abundances of the NGC\,6633 giants to those of the HeCB giants of similar masses in the Hyades, $\rm A(Li)_{LTE}$ = 0.86, 0.90, 0.99, and 1.11 \citep{Lambert1980}, and M48, $\rm A(Li)_{LTE}$ = 1.0 and 1.35 \citep{Sun2023}, shows that the Lithium abundances of NGC\,6633 giants are not unusual compared to other open clusters. The range of $\rm A(Li)_{LTE}$ = 1.34, 1.46, 1.37, 1.12, 1.27 \citep{Molenda2014} in the less massive 2.1-2.3 $M_{\odot}$ \citep{Arentoft2017} HeCB giants of NGC\,6811 serves to both confirm that the star-to-star A(Li) scatter is more than just observational errors and also that the average ZAMS rotational velocities must have been similar, since the difference in mean A(Li) compared to that of NGC\,6633 correspond roughly to what would be expected from the mass difference.    

\subsubsection{[C/N]}

\begin{figure}[ht!]
   \centering
    \includegraphics[width=\hsize]{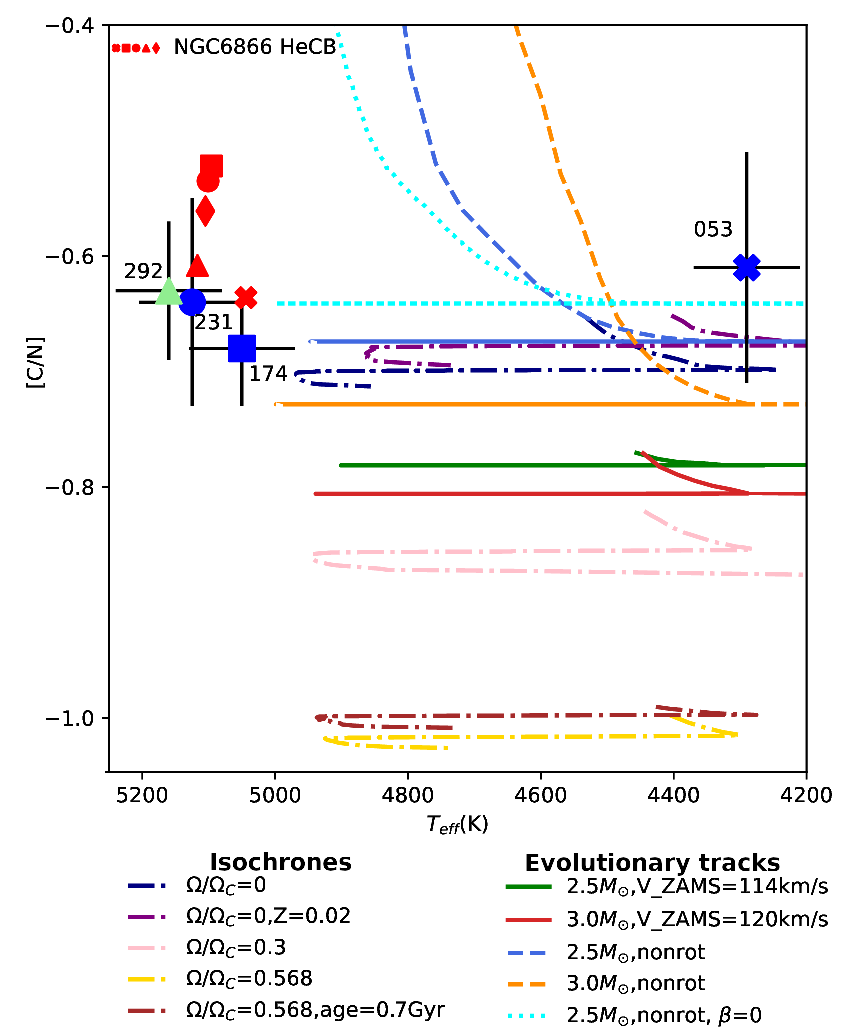}
      \caption{[C/N] measurements of HeCB stars as a function of $T_{\rm eff}$ compared to predictions from isochrones and stellar evolutionary tracks.
      Isochrones are from Syclist models of solar metallicity ($Z=0.014$) and an age of 0.5 Gyr unless otherwise stated. Evolutionary tracks of Solar metallicity are from \citet{Lagarde2012} with the exception of the 2.5 $M_{\odot},\rm{nonrot}, \beta=0$ track, which is from \citet{Vincenzo2021}.
      The different evolutionary stages are located similarly to Fig.~\ref{fig:Li}, but for clarity, many of the tracks and isochrones only show the part from the upper RGB to and including the AGB.}               
         \label{fig:CN}
   \end{figure}

In Fig.~\ref{fig:CN}, we compare the [C/N] measurements to the same stellar evolution tracks \citep{Lagarde2012} as for A(Li) in Fig.~\ref{fig:Li}. These are supplemented by a single evolutionary track without core-overshoot ($\beta=0$) from \citet{Vincenzo2021} and isochrones \footnote{Downloaded from https://www.unige.ch/sciences/astro/evolution/en/\\database/syclist/} from the Geneva Stellar Evolution Code 
\citet{Georgy2013, Ekstrom2012, Yusof2022}. Furthermore, we also show measurements of HeCB stars in NGC\,6866 (red symbols) from APOGEE DR17 \citep{APOGEEDR17-2022} as listed in \citet{Brogaard2023}.

Since the stars are in the stable HeCB evolutionary phase (with the exception of HD\,170053 which is either RGB or AGB), they should be compared to the hot end of the horizontal parts of the tracks and isochrones, which bend back to hotter temperatures after the RGB tip, as was the case for A(Li) in Fig~\ref{fig:Li}. As in Fig.~\ref{fig:Li}, either the effective temperatures of the models are too cool or the measurements are too hot, since the isochrones and measurements do not overlap in $T_{\rm eff}$. To avoid having a lot of overlapping lines, we show in most cases only the parts of tracks and isochrones from the upper RGB through the HeCB phase and the AGB.

\begin{figure}[h!]
   \centering
    \includegraphics[width=8.4cm]{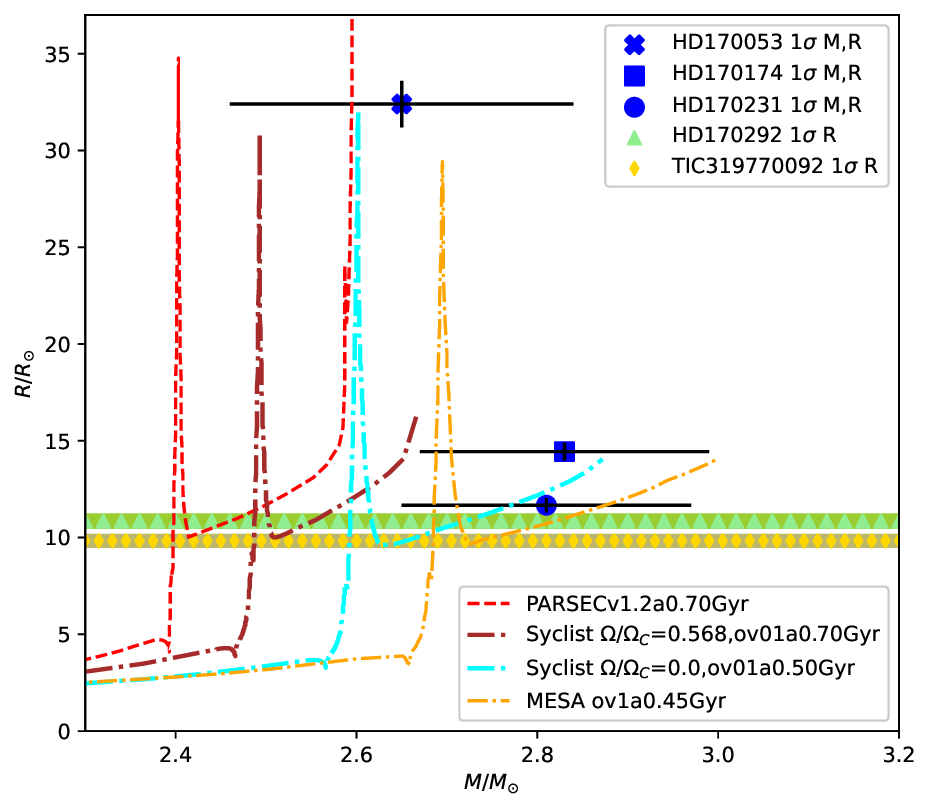}
    \caption{Mass-radius diagrams for NGC\,6633 giants compared to alternative isochrones with details in the text and legends.}
         \label{fig:mrt-rot}
   \end{figure}

The predicted dependence of $\rm [C/N]$ on mass, age, metallicity, rotation and core-overshoot can be estimated by inter-comparing different tracks and isochrones. For example, a comparison between the brown and yellow dash-dot isochrones close to $\rm [C/N]=-1.0$ shows that increasing the age by 0.2 Gyr reduces the depletion by about 0.02 dex. Similarly, comparing the blue dashed and green solid tracks shows that inclusion of rotation as predicted by the \citet{Lagarde2012} models increases the depletion by about 0.5 dex. Interestingly, comparing also to the  Syclist isochrones reveals a stronger dependence of [C/N] on rotation in the latter models, since the pink dash-dotted Syclist isochrone assumes $\Omega/\Omega_C=0.3$, which corresponds to $\rm V_{ZAMS}$ less than the 120 $\rm km\,s^{-1}$ of the red solid tracks, and yet it predicts a significantly stronger $\rm [C/N]$ depletion. If our asteroseismic mass measurements had been of higher precision, we could have used the mass-radius diagram to rule out this and the faster rotating models with $\Omega/\Omega_C=0.568$ as was done for NGC\,6866 in \citet{Brogaard2023}. However, due to the larger mass uncertainties in the present case, such a procedure only rules out the $\Omega/\Omega_C=0.568$ case, while allowing lower rotation rates as seen in Fig.~\ref{fig:mrt-rot}. Therefore, [C/N] and A(Li) are better indicators of rotational effects for NGC\,6633, at least until more precise mass measurements can be made for the giant stars. 

As seen in Fig.~\ref{fig:CN} by comparing to the observed values, even the non-rotating models seem to over-predict the depletion of [C/N] values at the HeCB phase, and the discrepancy increases significantly when including rotation in the models, even at a low level. It is worth noting that the relatively low star-to-star variation in $\rm [C/N]$ suggests only small star-to-star differences in the rotational velocities of the stars, even if interpreted as arising only from that. This is in agreement with our predictions from the A(Li) measurements above, though the suggested variation in $\rm V_{ZAMS}$ is even smaller than inferred from A(Li).
As can be inferred from the isochrones shown in Fig.~\ref{fig:CN}, realistic adjustments to metallicity and age within known constraints cannot solve the discrepancies. Comparing the measurements to those in NGC\,6866 implies that the stars in NGC\,6633 could be younger, less metal-rich and/or faster rotating on the ZAMS compared to NGC\,6866. Since age and metallicity is very similar for the two clusters and quite large differences are needed to explain the observations with these parameters, it seems more likely that a difference in ZAMS rotation speeds could be at play. However, the difference could also be partly caused by the fact that the measurements for the two clusters are from different instruments and wavelength regions (APOGEE for NGC\,6866 vs Gaia-ESO for NGC\,6633). Although we cannot quantify this, we note that both \citet{Spoo2022} and \citet{Casali2019} found significant changes for [C/N] values of cluster stars between different APOGEE releases. In any case, the [C/N] ratios are similar in the two clusters, suggesting that they have similar ages, though taken at face values, NGC\,6633 is suggested to be younger than NGC\,6866, at odds with all the other indications in this study.

The model evolution of [C/N] shown in Fig.~\ref{fig:CN} is not only due to rotational effects, but also other mixing processes like e.g. convective-core overshooting. To demonstrate the difference in their influence we include the cyan dotted isochrone in Fig.~\ref{fig:CN}, which shows a 2.5 $M_{\odot}$ track from the grid of \citet{Vincenzo2021} calculated without rotation and also without core-overshoot. This track is in better agreement with the [C/N] measurements, and could indicate that a lower value of $\beta$ is preferred. However, different stellar models were used to calculate this track, and a corresponding track with $\beta = 0.2$ (not shown) predicts the same [C/N] value for the HeCB and later phases as the \citet{Lagarde2012} track with the same mass, even though these models adopt $\beta = 0.1$. Therefore, half the difference in A(Li) between the dotted cyan and the dashed blue lines is due to other differences in the adopted model physics. This indicates a minimum error on the model prediction of [C/N] of about 0.02 dex even if mass and abundances could be measured without error. Therefore, [C/N] is a poor discriminator of core-overshoot within currently established range ($Beta=0.1-0.2$), but very useful for demonstrating that rotational effects on [C/N] are much smaller than predictions from current models.

Overall, both A(Li) and [C/N] suggest much more limited effects of rotation on the evolution of the giant stars in NGC\,6633 than predicted by the models. As mentioned earlier, $v\sin i$ measurements of bright main-sequence stars rule out the possibility that this is because $V_{\rm ZAMS}$ is lower than expected. If the initial A(Li) and [C/N] in the stars of NGC\,6633 were larger than assumed in the models, that could potentially reduce the tension, since then all tracks and isochrones in Figs.~\ref{fig:Li} and \ref{fig:CN} should be shifted upwards, bringing the rotating models closer to agreement with the observed values. However, this solution is also not likely; NGC\,6633 has near solar metallicity and \citet{Nieva2012} has demonstrated that a sample of nearby early B-type stars has abundances of Fe, C, N, and O quite similar to that of the Sun (see their Table 7). This suggests that the initial [C/N] of a star is not a function of age, though it might be a function of metallicity \citep[see e.g.][]{Vincenzo2021}. Therefore, it seems very unlikely that the initial [C/N] for NGC\,6633 was very different from that of the Sun. 

Regarding Lithium abundances, \citet{Jeffries2002} measured A(Li) for main-sequence stars in NGC\,6633, and in the $T_{\rm eff}$-range expected to be minimally affected by depletion on the main-sequence, they find A(Li) close to solar (A(Li)=3.0-3.2 depending on the temperature scale). Therefore, even if the pre-main-sequence Lithium abundance was significantly higher than solar, it would have been depleted to close to solar values on the main-sequence, removing the possibility to shift the tracks in Fig.~\ref{fig:Li} vertically.

We conclude from the above arguments that the giant stars in NGC\,6633 have been much less affected by rotation than predicted by the models, and that the main reason should be found in the physical description of rotational effects in the models rather than wrong assumptions on initial rotation speeds or initial abundances of Li, C and N. This is consistent with the findings of a low level of mixing for NGC\,6866 by \citet{Brogaard2023}. Similarly, \citet{GangLi2024} find that their isochrones prefer lower values of initial rotation than measured at the turn-off of the even younger open cluster NGC\,2516. 

\begin{table*}
\caption{Asteroseismic and other stellar properties of NGC\,6633 giant members}  
\label{tab:seispropdata}      
\centering                          
\begin{tabular}{l | c c c c c }        
\hline\hline                 
ID & HD\,170053 & HD\,170174 & HD\,170231 & HD\,170292 & TIC\,319770092 \\
\hline                                   
Symbol in plots & cross & square & circle & triangle & diamond  \\
$\nu_{\rm max}(\mu$Hz)\tablefootmark{a} & 8.99(18) & 44.75(10) & 67.47(19)& - & -  \\
$\Delta \nu (\mu$Hz)\tablefootmark{a} &  1.155(2) & 4.194(28) & 5.251(18) & - & -\\
$\Delta \rm{P_{obs}}$ (s)\tablefootmark{a} &  - & 237(13) & 170(12) & - & -\\
$f_{\Delta \nu}$ & 1.00 & 1.005 & 1.005 & -  & - \\
$R(\rm R_{\odot})$ seis, Eq. 1 & 34.32(08) & 14.19(22) & 13.76(15) &  - & -   \\
$R_{\rm SB}(R_{\odot})$ & 32.4(12) & 14.44(47) & 11.67(37) & 10.84(34) & 9.83(31)  \\
\hline                                   
$M(\rm M_{\odot})$ seis, Eq. 2 & 2.95(20) & 2.73(10) & 3.89(11)  &-&-\\
$M(\rm M_{\odot})$ seis, Eq. 3 & 2.51(29) & 2.89(29) & 2.39(23) &-&-\\
$M(\rm M_{\odot})$ seis, Eq. 4 & 2.65(19) & 2.83(16) & 2.81(16) &-&-\\
$M(\rm M_{\odot})$ seis, Eq. 5 & 2.86(14) & 2.76(06) & 3.52(04) &-&-\\
\hline
$<M(\rm M_{\odot})$ Eqs.>\tablefootmark{b} &  2.74 & 2.80 &  3.15 &-&- \\
rms$_{<M(\rm M_{\odot})\rm Eqs.>}$\tablefootmark{b} & 0.17 & 0.06 & 0.59 &-&- \\
\hline
$M(\rm M_{\odot})$ 'StarHorse2'\tablefootmark{c}  & $1.1_{-0.2}^{+0.4}$ & $2.7_{-0.4}^{+0.1}$ & $2.2_{-0.8}^{+0.2}$ & $1.5_{-0.3}^{+0.4}$ & $1.7_{-0.7}^{+0.4}$ \\
$M(\rm M_{\odot})$ Lagarde\tablefootmark{d}  & $4.2(9)$ & $2.7(6)$ & $3.4(6)$ & - &  - \\
$R(\rm R_{\odot})$ Lagarde\tablefootmark{d}  & $40.3(31)$ & $14.2(10)$ & $13.0(8)$ & - &  - \\
$\nu_{\rm max}(\mu$Hz)\tablefootmark{d} & 9.40(54) & 44.6(27) & 66.3(2.96)& - & -  \\
$\Delta \nu (\mu$Hz)\tablefootmark{d} &  1.09(3) & 4.15(8) & 5.34(11) & - & -\\
\hline                                   
\end{tabular}
\tablefoot{  
\tablefoottext{a}{This work.}
\tablefoottext{b}{Flat mean of the 4 equations for each star with asteroseismic measurements.}
\tablefoottext{c}{Starhorse2 catalogue of \citet{Starhorse2}.}
\tablefoottext{d}{\citet{Lagarde2015}}.
}
\end{table*}

\section{Comparisons to other mass- and cluster age estimates}
\label{sec:results}

We consider our age estimate for NGC\,6633 to be more precise and accurate than previous estimates in the literature because it is independent of uncertainties related to how the turn-off region of the CMD should be matched, and we have been able to demonstrate very low levels of extra-mixing, and thus that rotation has not had a significant influence on the age determination. Thus, it is currently more accurate to use non-rotating models for age estimation of NGC\,6633. With this in mind, we now compare our cluster age and HeCB mass estimates to other measurements from the literature.

\subsection{Comparisons to other age estimates}

NGC\,6633 is listed in the open cluster catalogues of \citet{Cantat2020} and \citet{Bossini2019} with $\log(\rm{age})$ = 8.84 and $\log(\rm{age})=8.888$ corresponding to 0.691 and 0.773 Gyr, respectively. These works are based on PARSEC v1.2 isochrones \citep{Bressan2012}, which were shown by \citet{Brogaard2023} to predict much larger radii for the beginning of the HeCB phase than PARSEC v2.0 \citep{Nguyen2022} and our MESA isochrones, and to be inconsistent with asteroseismic measurements of giant stars in NGC\,6866. In the case of NGC\,6633, the PARSEC v1.2 isochrones lead to similar discrepancy; the age would have to be significantly older than we derive to be able to match the smallest radii in the HeCB phase, but then the isochrone is not able to match the asteroseismic mass estimates, see Fig.~\ref{fig:mrt-rot}. The differences between these catalogue values and our work are therefore understood, and illustrate the need for more measurements of precise cluster properties to significantly improve catalogues of cluster ages. 

\subsection{Comparisons to other mass estimates}

The three stars with asteroseismology using CoRoT data were first measured by \citet{Morel2014} and masses from asteroseismic scaling relations (without corrections) were listed by \citet[][their Table 1]{Lagarde2015}. We repeat these in Table~\ref{tab:seispropdata}, where they can be compared to our measurements. They are close to our estimates using Eq. 1 and 2, since we derive the asteroseismic parameters from the same observational data. Their radii are not in agreement with the radii we derived from the luminosities using Gaia data, and their masses are only in agreement with our improved mass estimates because of their large uncertainties. We stress that the main improvement comes from utilising the luminosities, which is possible because of the Gaia DR3 parallaxes.

All our NGC\,6633 cluster member HeCB stars are listed in the StarHorse2 catalogue of \citet{Starhorse2}. We list the masses they derived in Table~\ref{tab:seispropdata} along with their $1\sigma$ uncertainties as suggested by their 16\% and 84\% quantile values. As seen, the masses are significantly lower than our estimates with one exception, and the uncertainties are in some cases suspiciously small for masses derived without using cluster membership information and asteroseismology. We suspect that these problems arise when trying to estimate ages from automatic isochrone fitting procedures for an evolutionary stage where systematic uncertainties in models can cause edge effects, since models may not cover the entire parameter space of the observed values and their uncertainties. For example, the temperature scale of the models could be too cool so that the observed temperatures and luminosities are only matched for a wrong mass. Similar discrepancies were seen between asteroseismic results and the StarHorse catalogue \citep{Queiroz2018} in the open cluster NGC\,6866 \citep{Brogaard2023}. Perhaps future versions of the StarHorse catalogue, and similar works, can begin to take advantage of such information.

\section{Summary, conclusions, and outlook}
\label{sec:conclusions}

We identify five giant members of NGC\,6633 according to similarities in their positions, proper motions, parallaxes, radial velocities, Gaia $v_{\text{broad}}$ values, metallicities, [C/N] and A(Li) values, effective temperatures, and magnitudes. We derive their luminosities and radii from the Stefan-Boltzmann equation. Combining this with asteroseismic measurements of three of the stars we improve previous estimates of masses and radii and obtain a self-consistent cluster age.
Our asteroseismic measurements allowed the evolutionary state of two of the stars to be established as the HeCB phase. Additionally, we classified the two giant members without asteroseismology as HeCB stars based on their similarities in colour, magnitude, radii, $T_{\rm eff}$, luminosity, $v_{\text{broad}}$ and by using time scale arguments for the different evolutionary phases as well as the agreement between age-inference from the mass-radius diagram and the brightest main-sequence star in the CMD. We were not able to establish the evolutionary state of the brighter giant HD\,170053, which could be either an RGB star close to the RGB tip, a very early HeCB star just after the RGB tip, or an AGB star, since the various observations provide only vague or inconclusive evidence. Establishing the evolutionary state of this star could help constrain the cluster age much more tightly in the future.
Comparing all the observations to stellar-model isochrones, we estimated an age of $0.55_{-0.10}^{+0.05}$ Gyr for NGC\,6633, including an uncertainty of 0.1 dex in metallicity and allowing for a convective-core overshoot parameter in the range $\beta=0.1-0.2$. [C/N] and A(Li) measurements from the literature were employed and suggest that rotational mixing effects are not significant for the evolution of the HeCB stars in NGC\,6633 and thus that non-rotating models are more appropriate for the best age estimate than current models including rotational mixing.

Our study could be improved significantly by obtaining asteroseismic measurements for all the giants members and with better precision and homogeneous measurements of $T_{\rm eff}$, A(Li), [C/N] and $v\sin i$ for all five cluster giants and for upper main-sequence stars. 

Comparisons to literature studies of the same stars in NGC\,6633 uncovered biases in mass and age. This underlines the importance of extending thorough combined asteroseismic-parallactic/photometric studies to more open clusters to improve models and methods for increased precision and accuracy of stellar age measurements. 
Unfortunately, the large pixel scale and the relatively short time span of observations in many regions hinders asteroseismology of solar-like oscillators in clusters using data from the TESS mission \citep{Ricker2014}. Therefore, although NGC\,6633 was observed by TESS in Sector 80 in July 2024, there is little hope that this will allow asteroseismic measurements of the two NGC\,6633 giants that have no measurements so far, and it will certainly not improve the asteroseismology for those already observed by CoRoT.

The large pixels will also reduce the quality of detailed asteroseismic studies in the crowded fields of clusters from the upcoming PLATO mission \citep{Rauer2014}, which is also not planned to observe the field of NGC\,6633. Instead, future missions like STEP\footnote{ https://space.au.dk/the-space-research-hub/step/} and Haydn \citep{Miglio2021b} will allow an extension of asteroseismic studies of cluster stars to a larger scale to ensure proper stellar ages of both cluster and field stars of all ages and metallicities in the future.

\begin{acknowledgements}
We thank Thierry Morel and Nadege Lagarde for clarifying questions and pointing to online data relating to the work of \citet{Morel2014} and \citet{Lagarde2015}.

We thank Laia Casamiquela for providing $T_{\rm eff}$ measurements of individual NGC6633 member giants from \citet{Casamiquela2021} on request.

Based in part on observations made with the Nordic Optical Telescope, owned in collaboration by the University of Turku and Aarhus University, and operated jointly by Aarhus University, the University of Turku and the University of Oslo, representing Denmark, Finland and Norway, the University of Iceland and Stockholm University at the Observatorio del Roque de los Muchachos, La Palma, Spain, of the Instituto de Astrofisica de Canarias.\\

This work has made use of data from the European Space Agency (ESA) mission {\textit{Gaia}} (\url{https://www.cosmos.esa.int/Gaia}), processed by the {\textit{Gaia}} Data Processing and Analysis Consortium (DPAC,\url{https://www.cosmos.esa.int/web/Gaia/dpac/consortium}). Funding for the DPAC has been provided by national institutions, in particular the institutions participating in the {\textit{Gaia}} Multilateral Agreement.\\
This research has made use of the SIMBAD database,
operated at CDS, Strasbourg, France\\
Funding for the Stellar Astrophysics Centre was provided by The Danish National Research Foundation (Grant agreement no.: DNRF106).\\
AM, KB, MT acknowledge support from the ERC Consolidator Grant funding scheme (project ASTEROCHRONOMETRY, \url{https://www.asterochronometry.eu}, G.A. n. 772293).

\end{acknowledgements}

\bibliographystyle{aa} 
\bibliography{References-1}

\begin{appendix}
\onecolumn

\section{Information on giant members of NGC\,6633}

%\Floatbarrier
\begin{table*}[h!]
\caption{Information on giant members of NGC\,6633}  
\label{tab:data}      
\centering                          
{\fontsize{9.4pt}{10.8pt}\selectfont
\begin{tabular}{l c c c c c }        
\hline\hline                 
ID & HD\,170053 & HD\,170174 & HD\,170231 & HD\,170292 & TIC\,319770092\\
\hline                                   
Symbol in plots & cross & square & circle & triangle & diamond  \\
Gaia DR3 ID & 447746039- & 447722337- & 447727326- & 447724911- & 447725330- \\
             & 1998886144 & 8527434880 & 8868842752 & 3972647168 & 5860760960 \\
2MASS ID\tablefootmark{a} & 18271429 & 18275474 & 18280018 & 18282297 & 18281763\\
         & +0700329 & +0636003 & +0654514 & +0642293 & +0646000 \\
RA (degrees)  & 276.809510  & 276.978078  & 277.000773   & 277.095725 & 277.073496 \\
Dec (degrees) & 7.009091 & 6.600086 & 6.914277  & 6.708119 & 6.766667 \\
pmra (mas$\cdot \rm yr^{-1}$) & 1.018 & 1.445 & 1.430 & 1.182 & 1.171\\
pmdec (mas$\cdot \rm yr^{-1}$) & -1.676 & -1.947 & -1.826 & -1.546 & -2.018 \\
Gaia DR3 parallax (mas) & 2.514(21) & 2.511(22) & 2.576(20) & 2.539(20) & 2.549(19)\\
$\nu$\_eff & 1.4085894 & 1.4590598 & 1.4565996 & 1.4639846 & 1.4669659\\
ecl\_lat (deg) & 30.263 & 29.84 & 30.158 & 29.947 & 30.007\\
RUWE & 0.962 & 0.958 & 1.013 & 0.879 & 1.035\\
Parallax corr. (mas) & -0.0329 & -0.0328 & -0.0326 & -0.0325 & -0.0324\\
$v_{\text{broad}}$ ($\rm km\,s^{-1}$) & $9.60(51)$ & $11.42(53)$ & $9.90(43)$ & $9.74(67)$ & $10.74(100)$ \\
RV ($\rm km\,s^{-1}$) & -29.27(13) & -28.71(13) & -28.45(13) & -29.02(14) & -28.68(16)\\
$G$ (mag) & 6.840 & 7.987 & 8.328 & 8.464 & 8.641  \\
$K_S$ (mag)\tablefootmark{a} & 4.087 & 5.674 & 6.065 & 6.246 & 6.443 \\
%$G-K_S$ (mag) & 2.753 & 2.313  & 2.263  &  2.218  & 2.198 \\
$G_{\rm BP}-G_{\rm RP}$ & 1.567 & 1.295 & 1.293 & 1.259 & 1.242  \\
$E(B-V)$ Green \tablefootmark{b} & 0.088 & 0.159 & 0.150 & 0.133 & 0.150 \\
$E(B-V)$ Pe{\~n}a \tablefootmark{c} & 0.182 & 0.182 & 0.182 & 0.182 & 0.182\\
$E(B-V)$ calc. $(V-K_S)$\tablefootmark{d} & 0.048 & 0.180 & 0.184 & 0.177 & 0.180 \\
$E(B-V)$ calc. $(G_{\rm BP}-G_{\rm RP})$\tablefootmark{d} & 0.112 & 0.185 & 0.205 & 0.189 & 0.187\\
\hline
$T_{\rm eff}$ (K)\tablefootmark{Ts} & -    & 4924 & 5015 & 5068 & 5093 \\
$T_{\rm eff}$ (K)\tablefootmark{So} & -    & 4979 & -    & 5124 & 5163 \\
$T_{\rm eff}$ (K)\tablefootmark{Ma} & -    & 4957 & 5025 & 5056 & - \\
$T_{\rm eff}$ (K)\tablefootmark{Mo} & 4290 & 5055 & 5175 & -    & - \\
$T_{\rm eff}$ (K)\tablefootmark{Ca} & -    & 5010 & 5095 & 5135 & 5165 \\
$T_{\rm eff}$ (K)\tablefootmark{Ad} & 4290 & 5050 & 5125 & 5160 & 5195 \\
$T_{\rm eff}$ from $G_{\rm BP}-G_{\rm RP}$ (K)\tablefootmark{e} & 4240 & 4953 & 4933 & 4970 & 5110 \\
$T_{\rm eff}$ from $G-K_S$ (K)\tablefootmark{e} & 4368 & 4987 & 5023  & 5029 & 5104  \\
BC$_G$ (mag)\tablefootmark{f} & -0.2603 & 0.0045 & 0.0194 & 0.0259 & 0.0321\\
BC$_{K_S}$ (mag)\tablefootmark{f} & 2.3773 & 1.8895 & 1.8447  & 1.8240 & 1.8033\\
$L(L_{\odot})$ from $K_S$\tablefootmark{f} & 321.3(53) & 122.4(21) & 84.7(13) & 75.0(12) & 63.35(93)\\
$R_{SB}(R_{\odot})$\tablefootmark{f} & 32.5(13) & 14.46(48) & 11.68(38) & 10.85(35) & 9.83(32) \\
\hline
$\rm [C/N]$\tablefootmark{Mo} & $-0.61$ & $-0.69$ & - & - & - \\
$\rm [C/N]$\tablefootmark{GE} & - & $-0.68$ & $-0.64$ & $-0.63$ & - \\
A(Li) NLTE\tablefootmark{MoMaTs}  & 1.44(12) & 0.80(12) & 1.49(12) & 1.21* & 1.42* \\
A(Li) LTE \tablefootmark{MoMa}  & 1.20(12) & 0.71(12) & 1.33(12) & 1.08* & - \\
A(Li) LTE \tablefootmark{Ts}  & - & 0.56(8) & 1.17(2) & 0.99(4) & 1.21(3) \\
A(Li) LTE\tablefootmark{MaMoTs} & 1.17*  & 0.66(5) & 1.31(3) & 1.05(3) & 1.28* \\
A(Li) NLTE\tablefootmark{MaMoWaTs} & 0.97(12)*  & 0.61(12) & 1.26(12) & 1.00(12) & 1.23(12)* \\
\hline
\end{tabular}
}
\tablefoot{  
\tablefoottext{a}{\citet{Skrutskie2006}.}
\tablefoottext{b}{From Bayestar2019 $E(g-r)$, using the conversion factor 0.884 from \citet{Green2019}, $A_G = 2.74\times E(B-V)$ \citep{Casagrande2018}, $A_{K_S} = 0.366\times E(B-V)$, $A_{G_{\rm BP}} = 3.374\times E(B-V)$ and $A_{G_{\rm RP}} = 2.035\times E(B-V)$ \citep{Casagrande2014}.}
\tablefoottext{c}{\citet{Pena2017}.} 
\tablefoottext{d}{Assuming adopted $T_{\rm eff}$s and requiring agreement with photometric temperatures from \citet{Casagrande2018}. The one using $V-K_S$ is adopted.}
\tablefoottext{Ts}{\citet{Tsantaki2023}.} 
\tablefoottext{So}{\citet{Santos2009}, values from "other linelist".} 
\tablefoottext{Ma}{\citet{Magrini2021}.} 
\tablefoottext{Mo}{\citet{Morel2014,Lagarde2015}.}
\tablefoottext{Ca}{\citet{Casamiquela2021}, numbers provided by L. Casamiquela (private comm.)}
\tablefoottext{Ad}{Adopted $T_{\rm eff}$s. See text.}
\tablefoottext{e}{Calculated using \citet{Casagrande2018} assuming $E(B-V)$ from \citet{Green2019}. [Fe/H] $=+0.0$ and log$g$ = 2.7 was assumed. A change of $\pm0.1$ dex in [Fe/H] corresponds to $\mp$8 K, a change of $\pm0.5$ in log$g$ yields $\mp2$ K} 
\tablefoottext{f}{Assuming the adopted $T_{\rm eff}$s, corresponding $E(B-V)$s, \citet{Casagrande2018}, and Gaia Parallaxes.}
\tablefoottext{GE}{\citet{GaiaESO1,GaiaESO2}}
\tablefoottext{MoLaMa}{A(Li) LTE from the online Table A5 of \citet{Morel2014}. For the measurements marked with a *, the abundance was calculated using the numbers from \citet{Magrini2021} and the mean difference for the stars in common.}
\tablefoottext{MoMaTs}{A(Li) NLTE from Table A5 of \citet{Morel2014} who used NLTE corrections from \citet{Lind2009}. For the measurements marked with a *, the abundance was calculated using the numbers of \citet{Magrini2021} or \citet{Tsantaki2023} and the mean difference for the stars in common.}
\tablefoottext{MaMoTs}{A(Li) LTE from \citet{Magrini2021}. For the measurements marked with a *, the abundance was calculated using the numbers of \citet{Morel2014,Lagarde2015} or \citet{Tsantaki2023} and the mean difference for the stars in common.}
\tablefoottext{Ts}{A(Li) LTE from \citet{Tsantaki2023}}
\tablefoottext{MaMoWaTs}{A(Li) 3D NLTE caluclated from \citet{Magrini2021} LTE values using 3D NLTE corrections from \citet{Wang2021}. For the measurements marked with a *, the 3D NLTE correction was applied to the LTE value of \citet{Morel2014,Lagarde2015} or \citet{Tsantaki2023} corrected to \citet{Magrini2021} using the mean difference for the stars in common. We adopt the larger of the error-bars from \citet{Morel2014} that includes systematic errors.}
}
\end{table*}
%\Floatbarrier

\section{More [C/N] discussion}

We elaborate here details of the [C/N] abundances that might affect conclusions.

We did not find any N measurements for NGC\,6633 dwarfs in the published literature, but \citet{Bertelli2018} did measure [C/H]. They found [C/H]$=+0.072\pm0.109$ from 5 main-sequence stars on the \citet{Grevesse2007} Solar scale and a reduction to [C/H]$=-0.200\pm0.044$ for 3 giants (that are among those of the present paper). While the measurements are unfortunately very uncertain, it could suggest that the models should have had a slightly higher initial C abundance and therefore shifted upwards by 0.07 dex in Fig.~\ref{fig:CN}, bringing the non-rotating models into agreement with the [C/N] measurements of the NGC\,6633 giants. However, without a higher precision measurement and a corresponding measurement of N, the increase of which contribute more to the [C/N] depletion than the decrease of C, we cannot know if this is the case.

The C and N measurements from the Gaia-ESO survey for the NGC\,6633 giants and from APOGEE DR17 for NGC\,6866 and the models used for comparison assume different abundances of C and N for the input Solar metallicity. This should, however, only have a very small impact on the comparisons. The difference between the Solar abundance of both C and N between the Solar abundance patterns of \citet{Grevesse2007} and \citet{Asplund2009} is 0.04-0.05 dex for both elements (the \citealt{Grevesse2007} values being smaller) with an additional 0.08-0.09 dex up to the values suggested by \citet{Grevesse1998} and further 0.03-0.05 dex to \citet{Grevesse1993}. Since the absolute abundances change by almost the same values between these works, the C/N ratio for the Sun is almost identical in all cases, and adopting one or the other does not matter for the zero-point of [C/N]. At a low level the assumed absolute abundances can make a difference for spectroscopic measurements as well as for the evolution of stellar models. Quantifying such assumptions is difficult, and we have not attempted to do so. However, most systematic effects should shift abundances of C and N in the same direction. We therefore find it reasonable to assume that the effects of such differences are smaller than those of the differences between models with and without rotation. On the model side, \citet{Lagarde2012} assumed the Solar abundances of \citet{Asplund2009} while the Syclist isochrones and the \citet{Vincenzo2021} models adopt those of \citet{Grevesse2007}, which are also those assumed for the spectroscopic measurements of Gaia-ESO, which we used as source for the [C/N] values. The fact that the \citet{Lagarde2012} and \citet{Vincenzo2021} models, which adopt different Solar abundances, predict identical [C/N] values for evolved giants of 2.5 $M_{\odot}$ with a difference in $\beta$ of 0.1 suggests that the assumed Solar abundance of the models does not affect the predicted [C/N] by more than 0.02 dex unless other model differences have a compensating effect (See the earlier discussion of core-overshoot). 

\section{Chemical age estimates of NGC\,6633}

\begin{table*}[h]
\caption{[Y/Mg]-age information for NGC\,6633}  
\label{tab:YMg}      
\centering                          
\begin{tabular}{l | c c c c c }        
\hline\hline                 
Source &  $\rm [Y/Mg]$ & $\rm [Y/Mg]_{M67}$ & $\rm [Y/Mg]_{corr}$ & $\rm [Y/Mg]-age$ (Gyr) & $\rm [Y/Mg]_{corr}-age$ (Gyr) \\
\hline                                   
\citet{Slumstrup2017} & - & +0.01 & - & - & - \\
\citet{Casali2020} & +0.08 & +0.00 & +0.07 & 2.4 & 2.7\\
\citet{Casamiquela2021} & +0.204 & -0.023 & +0.227 & < 0 & < 0 \\
\citet{Santrich2022} & +0.10 & -0.07 & +0.18 & 1.9 & 0.27 \\
\hline                                   
\end{tabular}
\end{table*}

\citet{Spoo2022} used [C/N] measurements in open clusters to derive a [C/N]-age relation for red clump stars. They include HeCB stars with masses larger than those corresponding to the secondary clump. This is not mentioned specifically, but their calibration clusters extend to young ages where this must be the case. We have listed the [C/N] values from the Gaia-ESO survey \citep{GaiaESO1,GaiaESO2} for our NGC\,6633 giant members in Table~\ref{tab:data} and used their mean value <[C/N]>$_{\rm NGC6633} = -0.64_{-0.05}^{+0.03}$ with the [C/N]-age relation in their Eq. 3 to obtain $\log(\rm{age}) = 8.72_{-0.11}^{+0.07}$, with the limits originating from the smallest and largest value of [C/N], respectively. This corresponds to an age of $0.52_{-0.12}^{+0.10}$ Gyr, which appear to be in excellent agreement with the age we derived earlier. However, one of the calibration clusters of \citet{Spoo2022} was NGC\,6866, for which one obtains a much larger age, although we have shown in this work that these two clusters must be quite close in age, and that NGC\,6866 is younger than NGC\,6633. 
\citet{Spoo2022} cut away clusters below $\log(\rm {age})=8.5$ from their calibration because they found a larger [C/N] scatter among clusters at very young ages. Perhaps our findings are showing that this scatter extends to higher ages. Currently, there is not a good explanation for the difference in [C/N] between these two clusters.

[Y/Mg] has been established to work as a chemical clock for giant stars of solar metallicity by \citet{Slumstrup2017}. We found measurements of [Y/Mg] for giants in NGC\,6633 and M\,67 in the works of \citet{Casali2020} and \citet{Casamiquela2021}.
In Table~\ref{tab:YMg} we list those values as well as the inferred age of NGC\,6633 from both the raw values, and those corrected to have the M\,67 [Y/Mg] values on the scale of \citet{Slumstrup2017} from which we took the age-[Y/Mg] relation. As seen, both statistical and systematic errors play significant roles in the limited precision of the inferred ages. Although the ages are consistent with the ones we derived using isochrones, the uncertainties are so large that all that can be concluded from [Y/Mg] is that NGC\,6633 is a relatively young open cluster. Determining [Y/Mg] from high-resolution high-S/N spectra using the exact same methods as in \citet{Slumstrup2017} could improve this, and the [Y/Mg]-age relation in general.

\end{appendix}

\end{document}